# Transparent and Electrically Switchable Thin Film Tactile Actuators Based on Molecular Orientation

Abigail Nolin[1], Chun-Yuan Lo[2], Laure V. Kayser[1,2], Charles B. Dhong[1,3]*

[1]Department of Materials Science & Engineering, University of Delaware, Newark, DE 19716, USA

[2]Department of Chemistry & Biochemistry, University of Delaware, Newark, DE 19716, USA

[3]Department of Biomedical Engineering, University of Delaware, Newark, DE 19716, USA

*Author to whom correspondence should be addressed: cdhong@udel.edu

**Abstract.** Most tactile actuators create tactile sensations through vibrations or the mechanical and electrochemical formation of bumps. However, tactile sensations of real objects arise from friction which is derived not only from physical topography, but also surface chemistry. Here, we show that molecular rearrangement can be leveraged to create new classes of tactile actuators based on the phases of liquid crystals embedded in a solid and transparent polymer film. We found that humans can feel differences by touch, especially between planar alignment and its disrupted phase, as actuated by a DC electrical field. In subjective terms, the sensation was described as a tacky to polished-like feeling. We attribute the mechanism of tactile contrast to microscale phase separation and changes in molecular orientation, as the nanoscale differences in topography are too small to be detected on their own by humans. This molecular rearrangement occurs quicker (<17 ms) than actuation through ionic or fluid movement. This enables a new class of tactile actuators based on molecular orientation (TAMO) for haptic interfaces.

**One-Sentence Summary:** We developed transparent and rapid tactile actuators based on the switching of molecular orientation which creates sensations that feel distinct from vibration or actuation of a bump.

## Introduction

Current tactile actuation is achieved by mechanical vibration (vibrotactile),[1–3] ultrasonic vibration,[4] electrical stimulation/adhesion,[5,6] or the actuation of bumps by either electromechanical or dielectric materials.[7–10] However, the current offering of tactile actuators are often slow, difficult to integrate on top of screens, and are restricted to a narrow variety of tactile sensations thus limiting adoption of haptics in virtual reality, remote surgery, and assistive aids for people who are blind.[11,12]

We previously showed that the chemistry and orientation of molecules can also create tactile sensations through the characteristic mechanical forces (i.e. friction) on a user's finger.[13–16] While novel polymeric materials with stimuli-responsive molecular properties have been explored as tactile actuators, these technologies ultimately produce physical bumps to alter mechanical forces, i.e., tactile stimuli on the user's finger.[17–19] In our approach here, we explored stimuli-responsive liquid crystal networks as a potential tactile actuator with changes in sensation derived solely from the inherent changes in molecular orientation and exposed chemistry, as opposed to actuation of roughness or formation of a bump. The molecular ordering of liquid crystal materials not only can influence bulk mechanics and viscoelastic adhesion phenomena,[20] but also surface adhesion shown in electrowetting applications.[21,22]

Here, we polymerized liquid crystals into solid films under three different molecular orientations. We characterized their surface mechanics and showed that human participants could feel different molecular orientation when actuated by heat or by an electric field. While stimuli-responsive materials have been previously used as macroscopic actuators in haptic applications, tactile actuators based on molecular orientation (TAMO) offer rapid actuation, are conducive to transparent and thin film fabrication, and generate distinctive tactile sensations from traditional tactile actuators.

## Results

**Fabrication of Polymer Network Stabilized Liquid Crystal Films**

Thermotropic liquid crystals can be fabricated into a solid polymeric network.[18,23–25] Here, liquid crystals were embedded into a liquid crystalline network. The polymer network stabilized liquid crystal (PSLC) films were prepared by mixing a high ratio (73 wt%) of nematic liquid crystal 4-cyano-4'-pentylbiphenyl (5CB) and liquid crystalline photoreactive monomer 1,4-bis[4-(3-acryloyloxypropoxy) benzoyloxy]-2-methylbenzene (RM257). In similar systems, a high LC:monomer mass ratio exceeding ~60% was necessary for switchable reorientation of the liquid crystal anchored between polymer grains, while having sufficient monomer to form a solid film.[22] This LC-monomer mixture is a nematic liquid mixture with a nematic-isotropic transition temperature ($T_{NI}$) at 56 °C (**Fig. S1**). To explore human sensitivity to molecular orientation, PSLC films of three different orientations were fabricated onto low roughness quartz wafers by using a corresponding alignment strategy (**Fig. 1A**). A mechanically buffed polyvinyl alcohol (PVA) alignment layer was used to achieve planar alignment (PA, homogenous alignment), where the average orientation of the liquid crystal molecules is parallel to the substrate. A hydrophobic octyltrichlorsilane (OTS) monolayer was used to achieve vertical alignment (VA, homeotropic alignment), where the average orientation of the liquid crystal molecules is perpendicular to the substrate. Finally, as a control, an unaligned film was crosslinked in its isotropic phase (Iso, disordered) which contains no director.

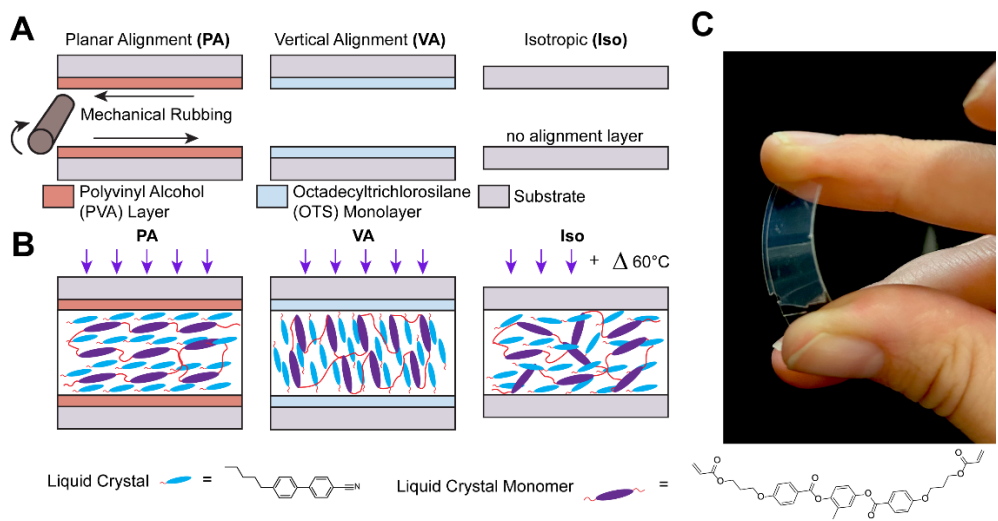

*Figure 1*: PSLC Film Fabrication. (A) Schematic of the three different surface alignment techniques to achieve films with different molecular orientations. (B) Schematic of polymerization conditions for synthesis of the three PSLC films. (C) Image of PA-aligned PSLC film supported on a transparent, ITO-coated PET substrate that demonstrates transparency and flexibility.

Surface treated wafers were pressed together to form a cell gap which was filled with the LC-monomer mixture through capillary action, and the top of the cell was exposed to UV light (**Fig. 1B**). After polymerization by UV radiation, the polymer network was able to stabilize the given liquid crystal orientation into a solid film. The reaction was carried out at room temperature for the aligned films to ensure the mixture remained in the desired orientation whereas the unaligned, isotropic films were heated above the $T_{NI}$. Polarized optical microscopy confirmed uniaxial alignment for the PA and VA films. (**Fig.**

**S2**) Both planar and vertically aligned films are visibly more transparent than the Iso film, and the transparency of the PA films were measured to be ~84% across the visible spectrum (**Fig. 1C** and **Fig. S3**)

## PSLC Film Characterization

**Grazing Incidence Wide Angle X-Ray Scattering (GIWAXS).** GIWAXS 2D patterns shown in **Fig. 2A** further confirms the molecular anisotropy and uniaxial alignment of the liquid crystal films. The pattern for the isotropic film is a diffuse, isotropic scattering ring indicating a molecularly disordered film lacking preferential orientation, while the patterns for PA and VA films show specific regions of scattering peak intensities indicating anisotropic alignment of the liquid crystal mesogens through π-π stacking.[26] For the VA films, the π-π stacking crystalline peak appears along the direction parallel to the substrate along the x-axis, indicating vertical alignment of the liquid crystal mesogens in the film network. For the PA film, the π-π stacking crystalline peak is seen perpendicular to the substrate, indicating planar alignment of the liquid crystal mesogens.[26]

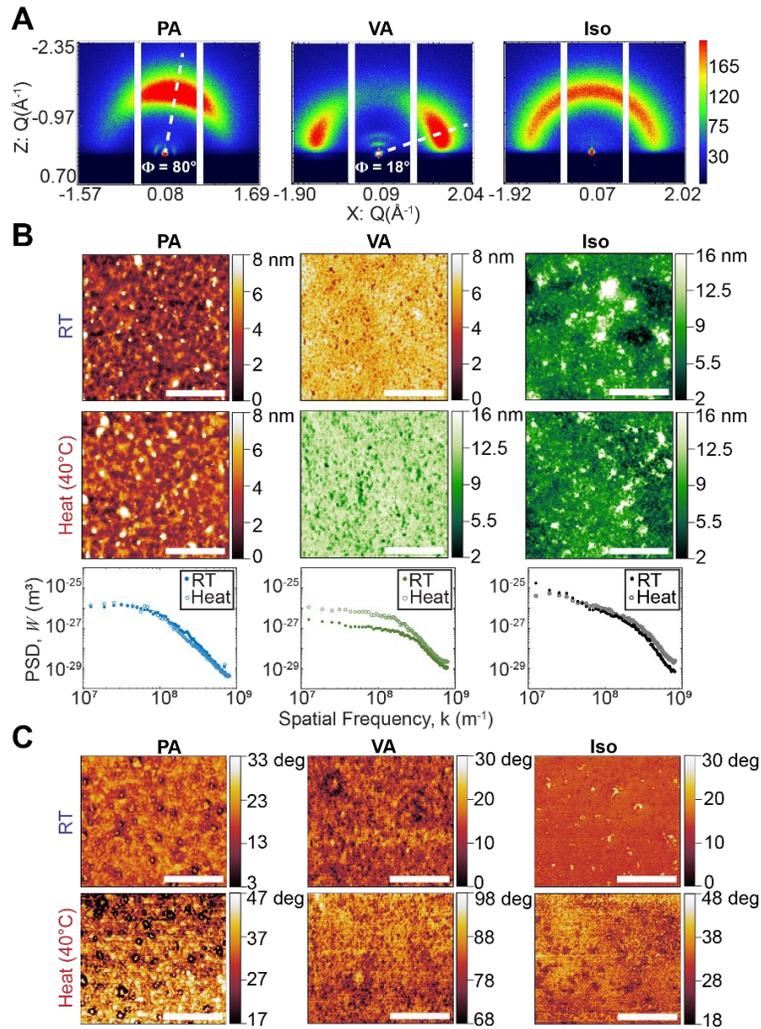

*Figure 2*: PSLC film characterization. (A) GIWAXS 2D patterns confirm differences in uniaxial alignment between the three films. Axes: scattering vectors. Scale bar: Intensity. The dotted white line represents the peak azimuthal angle (Φ). (B) AFM height images (1 μm x 1μm) indicating the surface topography and surface features. Scale Bar = 400 nm. Corresponding power spectrum density (PSD) analysis done by taking a Fourier transom of the height profile images above. (C) AFM phase images indicating different surface chemistry and material properties. Scale bar = 400 nm.

The degree of molecular alignment is quantified by the orientation order parameter ($S$), and obtained from the azimuthal intensity of the primary scattering peaks. The 2D patterns provide a corresponding 1D azimuthal intensity distribution, and $S$ can be obtained from integrating this intensity at the primary scattering peaks ($\Phi = 80°$ for PA and $\Phi = 18°$ for VA). $S = 0$ represents no preferred orientation or ordering and $S = 1$ represents perfect ordering. $S = 0.51$ was calculated for PA films, $S = 0.62$ for VA films, and $S$ was set to 0 in Iso films as an isotropic baseline.[26,27] These $S$ values are typical for aligned liquid crystal polymer networks retaining the nematic phase after polymerization.[9] Thus, the GIWAXS confirms that the aligned films retain different liquid crystal alignments after polymerization based on the surface alignment technique. GIWAXS also confirms the bulk Iso film, which was formed while heated, remains unaligned and molecularly disordered upon cooling to room temperature due to formation of the network.

Given that our TAMO design is based on the disruption of alignment, or molecular orientation, this liquid crystal ordering parameter $S$ defines an actuation range between the "on" and "off" state. Actuation of $S$ with heat showed successful disruption of anisotropic alignment in the PA film alignment by WAXS in **Fig. S4B**.

**Differential Scanning Calorimetry.** DSC heating and cooling curves shown in **Fig. S4A** display a small peak upon cooling and upon heating for all films at ~36-37°C. The bulk mixture of 5CB liquid crystal and liquid crystal monomer before polymerization had a $T_{NI} \sim 56°$ as confirmed by POM, however no phase transition appears at this temperature. The $T_{NI}$ for 5CB liquid crystal is ~37°C thus the DSC curves show that only the anchored liquid crystals are mobile enough to undergo the reversible phase transition from liquid crystalline phase (nematic) to disordered phase (isotropic), and not the polymerized network. This result does agree with literature in similar liquid crystal networks where the film is a polymerized solid,[18] and this high-density network precludes the material from undergoing thermotropic phase transitions. However, the overall LC network can undergo thermal expansions and strains (~5%) with heat and this is dependent on the crosslinking density as well as the temperature at which the films are polymerized.[18,28] In summary, DSC also demonstrates that the polymerized films retain reversible molecular reorientation from liquid crystalline to isotropic phases at ~37°C due to the anchored liquid crystal component.

**Atomic Force Microscopy (AFM).** To separate TAMO actuation from secondary actuation by topographical changes, we characterized the surface roughness of these PSLC films under AFM at room temperature and at 40° C (**Fig. 2B**). The average roughness parameter, $R_a$, of the PSLC films was calculated from the height images using Gwyddion software. All films had low roughness with an $R_a = 1.09$ nm for the PA film, $R_a = 0.579$ nm for the VA film, and an $R_a = 1.48$ nm for the Iso film. The differences in $R_a$ were all less than the perceivable limit to discriminate surfaces purely by roughness alone ($\Delta R_a > 7$ nm).[29] Upon heating, all films resulted in an increase in average roughness, with the VA film having the largest increase to an $R_a = 1.08$ nm, PA film with an $R_a = 1.14$ nm, and the Iso film with an $R_a = 1.69$ nm. Although the films become rougher with heat, these differences were all still below the perceivable limit,[29] thus discrimination between the films would likely not be driven by an apparent physical change.

During polymerization of aligned PSLC films, phase separation of the polymer network from the LC matrix occurs.[21,25] Both the PA and VA films show phase separated domains with regular spacing. The VA film has a rougher structure pointing outwards from the film, and smaller domain spacings compared to the PA film. However, the Iso film lacks regular domain spacing. Upon heating, both the PA and VA films show increased phase separation with larger domain spacings, with the VA film showing the largest changes in roughness and phase separation. These differences in characteristic dimensions of the surface can be quantified by the power spectrum density (PSD) of the height images on a log(PSD) versus log($k_{spatial\ frequency}$) plot shown in **Fig. 2B**.[30] Surface domains of regular spacing or different roughness are revealed by a peak or "knee" of two linear regions on the PSD plot. In contrast, randomly rough surfaces with self-affine characteristics will show a straight line with a constant slope. The spatial frequency is inversely related to the real space wavelength of surface features.[30–32] From the PSD plot in **Fig. 2B**, the PA film at room temperature shows a distinct intercept with a corresponding average domain size of ~50 nm. The VA film at room temperature shows a more pronounced intercept, and a much smaller corresponding domain size at ~30 nm. The isotropic film however lacks a regular domain spacing at room temperature. Upon heating, for both the PA film and especially the VA film, the intercepts of the PSD shifts to lower spatial frequencies, indicating larger characteristic domain sizes and phase separation.

The changes in the film surface features can be further realized with AFM phase images shown in **Fig. 2C**. The contrast differences in AFM phase images represent difference in material properties. All phase images were scaled to the same range around their individual mean phase value for clarity. The PA and VA films at room temperature show heterogeneity in phase across the surface. This is likely due to material property differences between the polymer network (RM257) and anchored liquid crystals (5CB). The isotropic film however does not show this same material heterogeneity as the aligned films, rather a relatively homogenous material surface with a low phase. This is likely because the network was polymerized while the liquid crystals were in an isotropic phase, and upon cooling did not undergo the same phase separation morphology as those polymerized in the nematic phase. Upon heating however, the isotropic film morphology becomes much more heterogenous and more similar to the PA and VA films under heat as seen by the contrast in the phase images.

The average phase for all films increased with heat, but this difference was most significant for the VA film increasing from an average phase of ~13° to ~ 83°. This suggests that the VA film network undergoes significant changes in surface morphology due to the alignment strategy, and this structural property change under heat is most pronounced in the VA films than in PA or Iso films.[18,24] Liquid crystal networks are able to undergo slight thermal expansions with heat, preferentially expanding perpendicular to the long axis of the molecule, inducing phase separation. In the PSLC films here, this manifests as exposing liquid crystalline domains at the surface and the magnitude of exposure of these domains depends on the alignment strategy.

In conclusion, surface characterization reveals that the PSLC films have differences not only in liquid crystalline alignment confirmed by WAXS and POM, but also different surface morphologies which are influenced by heat, especially for the VA film. All films have two components: one, a liquid crystalline component which is anchored at a different orientation depending on the alignment strategy and can be disrupted with heat, and second, a network component which influences the surface morphology and phase separations and differs depending on the surface alignment technique. While the network itself does not exhibit a phase transition with heat, the bulk film with a high fraction of liquid crystals undergoes a thermal transition which changes the phase separation within the film and the moieties exposed to the surface. The degree of phase separation and changes under heat in the composite films are unique to the underlying alignment of the liquid crystals.

**Mechanical Testing.** As we will show here, it is possible for two surfaces which exhibit larger differences in material properties or topography to have comparatively smaller differences in friction.[13] To rationally determine which surfaces would generate the largest tactile actuation between the "on" and "off" state, we measured mesoscale friction between the films and a "mock" finger at both at both room temperature and at 40 °C with a custom mechanical setup shown in **Fig. 3A**. A similar mesoscale friction setup and method was used in our previous studies.[13,14,16]

The mock finger was slid across each surface at 16 different sliding conditions (4 velocities and 4 applied masses) relevant to human touch exploration[13] to analyze the unique stick-slip friction present in the friction traces. Stick slip friction is due to transient trapping of the soft finger (stick) and subsequent energetic releases (slip) on the mesoscale, and is dependent on the applied mass, velocity, and orientation of the films, as actuated by temperature. **Fig. 3B** shows representative friction traces and differences in stick-slip friction amongst the differently aligned PSLC films. The PA film shows the highest magnitude stiction spike followed by higher amplitude and slower oscillations compared to the VA and Iso films. However, this stick-slip friction behavior changes upon heating, with all films reducing in stiction, but significantly for the PA film. The different oscillations seen in the stick slip friction traces are at amplitudes perceivable to touch ($\mu = F/F_n > 0.035$).[33]

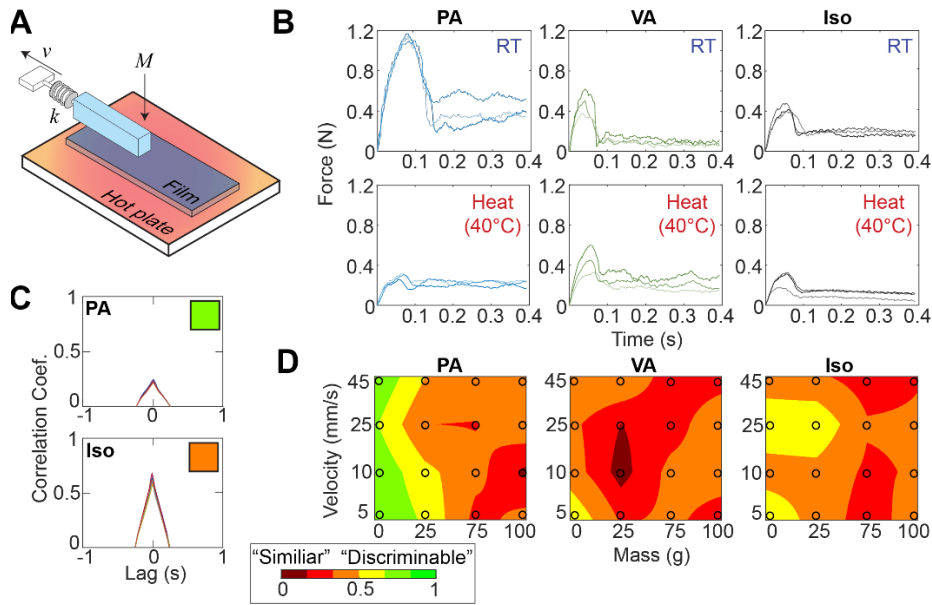

*Fig. 3*: Mechanical testing and cross-correlation analysis. (A) Schematic of mechanical testing set-up. An elastic mock finger is brought into 1 cm x 1 cm contact and slid tangentially across the films at either room temperature or heated to 40°C at a given sliding velocity, $v$, and applied mass, $M$, onto the finger. The collected force is transduced by a force sensor with spring constant, $k$, to collect sliding frictional force on the mock finger. (B) Representative friction traces from all three PSLC films collected at room temperature (RT) and at 40°C at the same condition where $v = 10$ mm/s and $M = 0$ g. (C) Cross-correlation analysis of two different surfaces, PA and Iso, at the same condition ($v = 10$ mm/s and $M = 0$ g), quantifying the similarity of the friction traces for each material at RT versus at 40°C. (D) Discriminability matrices summarizing the differences in friction at RT versus with heat for all three films. In this case, differences are quantified by the average correlation of the cross-correlation curves. Black circles represent the conditions tested, and a heat map is generated by 2D interpolation. All friction was measured in triplicate and repeated on three different spots.

At room temperature, PA tended to show a more adhesive surface and produced the largest magnitude stiction spike at most conditions. In general, large stiction spikes indicate an adhesive contact, and have been seen in sliding friction of nematic liquid crystal materials.[17] In contrast, liquid crystals in isotropic phase have shown to reduce stiction, or static friction, compared to their more adhesive nematic state,[20] as we also see here.

To synthesize friction from multiple experimental conditions into a predictor of human performance, we quantified the similarity or dissimilarity of friction traces from each film through cross-correlation (**Fig.**

**3C)**, which we previously showed is better correlated with human results than using friction coefficients.[13] A cross-correlation versus lag is calculated by **Eqn. 1**:

$$Cross-Correlation = \frac{\sum(a(t) - \hat{a})(b(t) - \hat{b})}{\sqrt{\sum(a(t) - \hat{a})^2 \sum b(t) - \hat{b})^2}} \quad (1)$$

Where *a* and *b* are the time-series of the two friction traces being compared and the hat signifies mean value. Similar friction traces generate high correlation (large, symmetric in **Fig. 3C**) and are likely harder for humans to discriminate by touch whereas distinctive friction traces that generate low correlation (low, asymmetric in **Fig. 3C**) are likely easier for humans to discriminate. These cross-correlation versus lag vectors can be further parameterized into a single value by an average cross-correlation (with red representing high averages, difficult for humans to distinguish, and green representing low averages, easier for humans to distinguish) or some other parameter. This analysis condenses all the stick slip friction traces from a comparison of two surfaces or conditions into a single value on the scale of "similar" to "discriminable" (red-to-green scale), and plotted for every mass and velocity tested forms a "discriminability matrix."[13,14,16]

In **Fig. 3D**, discriminability matrixes were constructed of the same film orientation but compared at room temperature versus heat. This determines which PSLC orientation would produce a TAMO actuator with the strongest tactile difference between their "on" and "off" state. In **Fig. 3D,** the discriminability matrix suggest that PA orientation undergoes the largest changes in mesoscale friction with heat. Discriminability matrices were also constructed with the average skew parameter (**Fig. S5**) from cross-correlation traces and suggested the PA orientation underwent the largest changes in mesoscale friction. Interestingly, this film showed the smallest changes in roughness and phase as characterized in AFM.

**Human Testing.** To verify TAMO actuation of the PSLC films, we performed human participant studies with both heat-actuation and electrical-actuation to establish TAMO actuation mechanisms and end-device demonstration, respectively. Data were collected from a total of 14 healthy volunteers between the ages of 18 and 40 (see acknowledgements for IRB approval).

**Heat Actuation.** For heat actuation, we could not simply compare a single film unheated versus that film heated, as subjects would know which film was different based on temperature. Instead, we used a three-alternative forced choice task, or "odd-man out" task that compared two different PSLC orientations once at room temperature and then again with heat. We then quantified how human accuracy changes between room temperature versus heat. The testing setup and combinatorial results of the three different PSLC films are shown in **Fig. 4A and Fig. 4B**. Each participant performed each comparison for 5 trials. The first set of comparisons was performed at room temperature conditions. As the Iso film has a different visual appearance from the VA and PA films, subjects were blindfolded for all the room temperature comparisons. In the heated case, films were heated to 40°C on a hot plate. For these experiments, participants vision was occluded by a black box, but participants were not blindfolded as a safety precaution. However, as shown in **Fig. 4A**, at 40°C, all film orientations became visually identical.

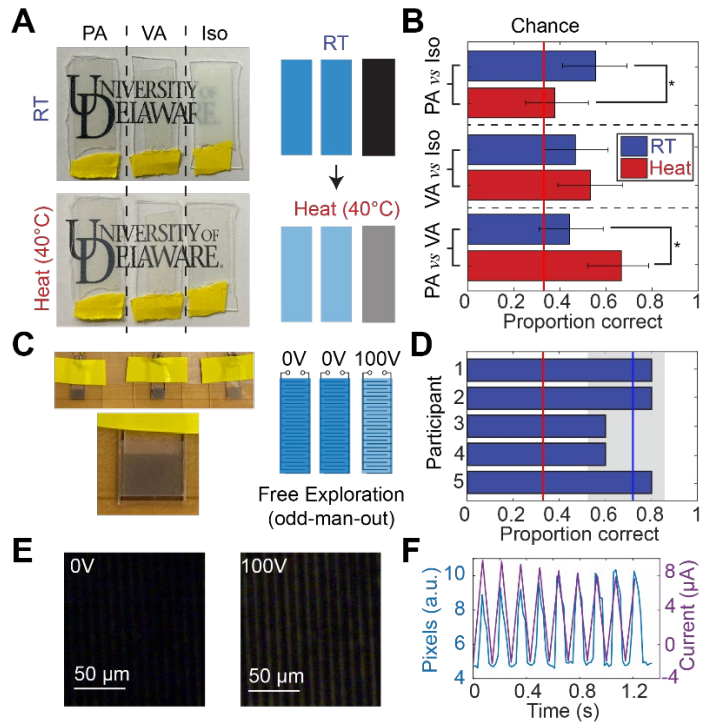

*Fig. 4*: Human psychophysical testing. (A) Heat actuation. Images of the three films at room temperature (RT) and on a hot plate heated to 40°C. Schematic of "Odd-man-out" (three alternative forced choice task) for the comparison of two differently aligned films at RT and at 40°C. (B) Results from human participants. Each comparison had 45 trials with n = 9 subjects. (*) indicates p < 0.05. (C) Electrical actuation. Images taken of three PA films fabricated onto gold patterned electrodes. Schematic of "Odd-man-out" setup. (D) Results from human participants. Task was performed for 25 trials with n = 5 subjects. 95% confidence interval calculated by Wilson Test. The blue line represents the average response. (E) Crossed POM images of the PA film on electrode with no applied electrical field (0V) and at an applied voltage of 100 V. At 0V, the PSLC film is aligned with the electrodes and the axis of the polarizer. At, 100V, light transmission increases due to the molecular reorientation in the film induced by the electric field. (F) Actuator response time measured by plotting the applied signal against the responding pixel brightness collected in a video recording on the POM camera.

Human testing results in **Fig. 4B** show that VA and PA, although completely opposite orientations, were slightly but not significantly distinctive through touch (accuracy = 44.4%, z = 1.47, p > 0.05). However, VA versus Iso films were slightly more distinctive at room temperature (accuracy = 46.7%, z = 1.791, p < 0.05) and the PA films versus the iso films were significantly distinctive through touch (accuracy = 55.6%, z = 3.06, p < 0.01). These results indicated that opposite alignments, planar versus vertical, had less of an influence on tactile perception compared to an aligned versus isotropic alignment.

All three comparisons were then repeated at 40 °C. PA versus Iso films under heat became significantly less distinctive under heat and very close to chance (accuracy = 37.8%, z = 0.523, *p* > 0.05). This difference in perception at room temperature versus under heat was statistically significant determined by a z-score test of 2 population proportions (z = 1.69, *p* < 0.05). However, VA versus Iso films under heat became slightly *more* distinctive under heat (accuracy = 53.3%, z = 2.742, *p* < 0.05), although not significantly (z = 0.6325, *p* > 0.05). This was an opposite trend compared to the PA versus Iso film comparison. Interestingly, when comparing PA versus VA films under heat, although now both less ordered with their characteristic alignment disrupted, and hypothesized to become more similar, became significantly distinctive (accuracy = 66.7%, $z = 4.64$, $p < 0.01$).

This result is evidence that simply disruption in orientation (order parameter) is likely not the only predictor of fine touch distinction as PA and VA films had higher differences in alignments at room temperature compared to their heated state, yet to humans, PA and VA became *more* distinctive with heat. Rather, the overall network structure formed during synthesis due to alignment method and polymerization conditions altered the change in surface morphology of the materials under heat, with different material properties exposed as the bulk film was disordered. The VA network did indeed undergo the largest network change in terms of topography and phase shown in AFM.

**Electrical Actuation.** Human testing results as well as mesoscale friction suggested that the PA films were a good candidate for a TAMO device demonstration. Shown in **Fig. 4C**, films with PA orientation were fabricated onto comb gold electrodes, with the alignment director parallel to the electrodes as verified with POM. For human testing, we again employed a three-alternative forced choice task, but this time the participant was presented three of the same PA films, with one of the films under electrical field at 100V and a current of ~ 6 - 8μA. Upon applying a DC electrical field across the film, the PSLC film switches from a dark state to a bright state with light scattering in a pattern based on the comb electrode dimensions due to molecular reorientation in the film.[22] A DC voltage of 100V was sufficient to cause molecular reorientation of the film, as visible with cross polarizers (**Fig. 4E**).

As before, the location and the alternate were randomly selected every trial. Each participant completed the task for 5 trials. The average accuracy for identifying the "on" from "off" PA film was 72%, and significantly above chance ($p < 0.01$), where all participants were able to discriminate the surface above chance (**Fig. 4D**). Participants described the difference as changing from a "tacky-like" feel (voltage off) to a "smoother, polished-like" feel (voltage on). This human participant studies demonstrated that the PSLC films in planar alignment can effectively actuate fine touch with electrical field through molecular orientation. Interestingly, despite molecular ordering being only reorientated between the electrodes across the film, humans had a higher accuracy on electrically-actuated PA films than under heat, where the entire film becomes molecularly disordered.

This increase in accuracy for electrical actuation may have been because several reasons: One, with heat actuation, as it was applied underneath the entire film, network morphology changed much more with heat (and to different magnitudes depending on the film alignment), masking the influence of molecular reorientation on actuation. Second, heat stimuli could be distracting regardless of texture and attenuate other tactile stimuli.[34] Third, the increase in accuracy may have been due to the underlying patterning of the electrode substrate, and as a result, a patterning of friction which was not present in the heated films. Taken together, in the design of TAMO actuators, parameters to be considered are not only controlling the orientation and order parameter of the mobile molecules, but also the phase separation of the network morphology to control the proportion of molecules exposed at the surface and the spacing between them.

To quantify the actuator response time of the PSLC film, an electrical field was generated through an electrical bias of 100V as applied in a repeated cycle with a driving frequency of 7 Hz (1 cycle every ~0.14 s), and the film was recorded under crossed polarized microscopy to measure the speed of actuation through pixel brightness (**Movie S1**). **Fig. 4F** shows that the film switched from a dark state to a bright state with a rapid response to electrical stimuli. The difference in peaks times between the current signal and pixel brightness were calculated to be lower than the recorded frame rate of the camera, 1 frame/17 ms or ~ 60 frames/s. Therefore, the response time was limited to the camera frame rate and is faster than ~ 60 Hz. This fast response is also evidence that the mechanism of switching is molecular reorientation to the electric field, not based on joule heating. The mechanical properties of the PSLC likely facilitated this rapid response through low viscoelasticity.[18]

## Discussion

Here, we showed a tactile actuator based on molecular orientation (TAMO), in contrast to most tactile actuators based on bulk physical motion or the formation of bumps. The actuator qualitatively changed from a tactile feel of "tacky" (off) to "polished" (on), which generates tactile sensations distinct from those made by vibrotactiles or by actuating bumps, regardless of whether the bump is formed by traditional electromechanical actuators or by modern soft actuators. By using liquid crystals embedded into a network film, we were able to achieve actuation at DC voltages below those used in dielectric soft actuators (DEA) and speeds much faster than electroactive polymers (EAP).[34,35] Adoption of our tactile actuator into existing technologies like screens and curvilinear substrates is facilitated by both its transparency, being cast from solution, and the fact that actuation is achieved by a DC electric field achieved in a single device layer, as opposed to direct bias across an active layer.

We used heat to systematically disrupt the order in different film orientations and established that the mechanism of TAMO can be attributed to both the liquid crystal orientation and the network morphology. We found that in comparing the PA *vs* Iso films, application of heat decreased accuracy, which was attributed to the disruption of the PA nematic phase into a more isotropic phase, thus resulting in a film that is more difficult to tell apart from the Iso film due to liquid crystal orientation. However, in comparing VA *vs* Iso films, application of heat *increased* distinction. Atomic force microscopy showed that the topography of the film changed the most for VA morphology, thus we can attribute increased accuracy under heat due to significant changes in the VA network morphology changing the phase separation and surface composition, overriding the loss of orientation in the liquid crystals as the sole drive of actuation.[36] Finally, we excluded that the primary mechanism is by changing roughness because the roughness changes by AFM in these films was on the order of $\Delta R_a < 1$ nm, and because the degree of change in roughness did not correlate to higher human accuracy in distinguishing surfaces. Characterization of mesoscale friction measurements, however, did correlate to higher accuracy, and thus should be included in the characterization of these actuators.

Finally, we found that electrical actuation of a single alignment (PA) provided higher accuracy than any heated films, even though the heated films undergo complete loss of their alignment across the entire film. We attributed this high performance in the electrical actuation to the pattern generated by the underlying comb electrode: the PA film actuates in the regions between electrodes, thus creating alternating stripes of molecular orientation in the PA film. It is known that patterning can accentuate differences in friction.[37] Thus, electrode design, in addition to liquid crystal network formulation, could be further explored for TAMO.

## Materials and Methods

### Fabrication of PSLC films

The polymer network stabilized liquid crystal (PSLC) films were prepared by first mixing nematic liquid crystal 4-cyano-4'-pentylbiphenyl (5CB, TCI Chemicals) and liquid crystalline photoreactive monomer 1,4-bis[4-(3-acryloyloxypropoxy) benzoyloxy]-2-methylbenzene (RM257, TCI Chemicals) at a 5CB: RM257 ratio of 73 wt. %. Materials were mixed at 80°C until solution became clear, producing a homogenous, nematic liquid mixture with a nematic-isotropic Transition Temperature ($T_{NI}$) at 56°C.

For planar alignment, the quartz wafers were spincast with 7% PVA (purchased from sigma Aldrich) solution at 2300 rpm for 45 seconds and then baked at 120°C for 2 hours. PVA surfaces were then mechanically buffed with a microfiber cloth in anti-parallel directions. For vertical alignment, quartz wafers were exposed to air plasma (Glow Plasma System, Glow Research) for one minute to introduce reactive hydroxyl surfaces onto the surface. Wafers were immediately transferred to vacuum desiccators containing ~50 uL of octyltrichlorosilane (OTS) on a glass slide. Desiccators were than evacuated and held under static vacuum for 48 hours at 90°C. For the isotropic film, untreated quartz glass was used as the underlying substrates.

Surface treated wafers were pressed together to form a cell gap which was filled with the LC-monomer homogenous mixture through capillary action. LC-monomer containing cells were heat cycled at 65°C for 1 minute to erase flow marks and then cooled back to room temperature and a nematic phase state. Once alignment was formed and confirmed by crossed polarizers, the cell was exposed to UV light (365 nm) with intensity $I$ = 10 mW/cm$^2$ for 30 minutes. The reaction was done at room temperature for the planar aligned (PA) nematic sample and vertical aligned (VA) nematic sample, and a curing temperature of 65°C for the isotropic sample. The top glass substrate was removed via a thermal release process and a solidified polymer network stabilized liquid crystal (PSLC) film with a thickness ~ 45 um remained. The PSLC film was rinsed with DI water to remove any unanchored liquid crystal and left to dry.

**Fabrication of Electrical PSLC films**

For electrically responsive configuration, PSLC films were fabricated onto glass substrates overcoated with an interdigitated, comb gold electrode pattern with a width and gap of 5 μm (purchased from Metrohm). The electrode surface was used as the bottom substrate in fabrication. The top substrate was quartz glass spincast with a thin PVA buffed polymer layer rubbed along the same direction as the electrode stripes. As actuation of the PSLC film occurs from the presence of an electric field, direct electrical contact between the film and electrode is not necessary. Afterwards, the cell gap was filled with LC/monomer mixture and exposed to UV light on the side of the top substrate for 30 min at room temperature. Successful planar alignment was checked with crossed polarized optical microscopy.

**Surface Characterization**

**Grazing Incidence Wide Angle X-Ray Scattering.** (GIWAXS, Xenocs SAXS/WAXS) 2D scattering patterns were obtained at room temperature and with a wavelength = 0.154 nm and incident angle of 0.2°. GIWAXS scattering profiles were used to confirm uniaxial alignment and quantify the orientation order parameter (S) of the aligned films. Wide Angle X-ray scattering was also collected on the PA film in the upright orientation module where the beam penetrated through the entire film. 2D patterns were collected under this confirmation at both room temperature and under heat to confirm the disruption of anisotropic molecular ordering.

**Differential Scanning Calorimetry.** (DSC, TA instruments) experiments were performed on TA Instrument Discovery series to characterize the phase transition temperatures and properties (reversible nematic to isotropic transition) of the final PSLC network films after fabrication. All films were tested at a heating rate of 10 °C/min from -80 to 80 °C under N2 atmosphere and observed for crystallization and melt peaks.

**Atomic Force Microscopy.** (AFM, Bruker Multimode, analyzed with Gwyddion software) Height and phase AFM images were collected on all three alignments (PA, VA, Iso) at both room temperature and at 40 °C using a temperature module powered by a NanoScope 6 controller. AFM height images were collected to characterize surface roughness and topography, and phase images were collected to characterize material chemistry and composition present at the surface. Surface images were obtained through tapping mode over a scar are of 1 μm x 1 μm area at a scan rate of 1 Hz and drive frequency of 300 kHz with RTESPA-300 Burker AFM tips.

**Mock Finger Preparation.** The mock finger was prepared with the same methods used in our previous work.[13,14,16] The mock finger was prepared by curing polydimethylsiloxane (PDMS) in a mold with dimensions of 1 cm x 1 cm x 5 cm around a resin 3-D printed 'bone" at 60°C for one hour. The dead weight of the mock finger was 5.5 g. The elastic modulus of the PDMS was controlled by the ratio of base to crosslinker, which was 30:1 to provide a modulus of ~100 kPa. The 3D printed bone provides mechanical rigidity and stiffness, and the PDMS provides a low modulus, achieving an effective modulus similar to real fingertips. As real fingers are deformable, but not sticky, the cured PDMS slab was treated with UV/Ozone for four hours to remove the viscoelastic tack at the surface to achieve a surface energy similar to human skin (60° water contact angle). The mock finger controls for all the critical material properties of a real human finger to capture the frictional forces present in tactile exploration.

**Mechanical Testing.**

The same mechanical testing set-up for this work has been used in our previous work to collect stick-slip friction traces for predictions of human tactile performance. To collect friction of the mock finger when sliding against the PSLC films, the finger was loaded with masses, $M$ = 0, 25, 75, or 100 g, then brought into a contact length of 1 cm x 1 cm with the film, and slid for a distance of 4 mm at velocities, $v$ = 5, 10, 25, or 45 mm s$^{-1}$. The finger was slid using a linear motorized stage (V-508 PIMag Precision Linear stage, Physikinstumente) while friction force was measured with a Futek 250 g forces sensor ($k$ = 13.9 kN m$^{-1}$, peak-to-peak noise of 0.1 mN) screwed into the mock finger. The sampling rate was 800 Hz (Keithley 7510 DMM) to sufficiently capture mechanical signals relevant to mechanoreceptors in fine touch (~40 – 400 Hz).[38] For every collection, the finger was always slid four times, where the first slide was discarded to eliminate artifacts due to surface aging. Therefore, three slides on three different locations of the film were collected and analyzed for each condition (16 combinations of mass and velocities), totaling nine friction traces for each condition.

**Acknowledgements**

**Funding**: National Eye Institute of the NIH (R01EY032584). L.V.K. acknowledges funding from the Arnold and Mabel Beckman Foundation through a Beckman Young Investigator award. Atomic force microscopy access supported by grants from NIH-NIGMS (P20 GM103446), NIGMS (P20 GM139760) and State of Delaware.

We would like to acknowledge Katie Herbert, the scientific program manager at the Center for Platics Innovation at the University of Delaware, for her advice on liquid crystal fabrication. This study was conducted and approved by the Institutional Review Board (IRB) of the University of Delaware (Project #1484385-2).

AN and CBD declare a potential conflict of interest as they are seeking a patent for this work.

**Author contributions:**
Conceptualization, Investigation – AN, CBD
Methodology – AN, CYL, LVK
Funding acquisition – CBD, LVK
Supervision – CBD, LVK
Writing – AN, CBD

Editing - LVK

**References**.